\documentclass{svproc}
\usepackage{url}
\usepackage{amsfonts}

\def\>{\rangle}
\def\<{\langle}
\newcommand\hull{\mathrm{hull}}

\begin{document}
\mainmatter              
\title{Non-additive measures for quantum probability?}
\titlerunning{Non-additive measures for quantum probability?}  
%
\author{Gabriele Carcassi \and Christine A. Aidala}
\authorrunning{Gabriele Carcassi and Christine A. Aidala} 
%
%
\institute{University of Michigan, Ann Arbor, MI 48109, USA\\
\email{carcassi@umich.edu}
}

\maketitle              

\begin{abstract}
	It is well-established that quantum probability does not follow classical Kolmogorov probability calculus. Various approaches have been developed to loosen the axioms, of which the use of signed measures is the most successful (e.g. the Wigner quasiprobability distribution). As part of our larger effort Assumptions of Physics, we have been considering the various roles of measures, which are used in physics not only for probability, but also to quantify the count of possible states and configurations. These measures play a crucial role in classical mechanics, as they effectively define its geometric structure. If one tries to construct a parallel in quantum mechanics, the measure to quantify the count of states turns out to be non-additive. The idea, then, is that the proper extension of probability calculus may require the use of non-additive measures, which is something that, to our knowledge, has not yet been explored.
	
	The purpose of this paper is to present the general idea and the open problems to an audience that is knowledgeable of the subject of non-additive set functions, though not necessarily in quantum physics, in the hope that it will spark helpful discussions. We will go through the motivation in simple terms, which stems from the link between the entropy $S$ of a uniform distribution $\rho_U$ and the logarithm of the measure $\mu$ associated to its support $U$ (i.e. $S(\rho_U) = \log (\mu(U))$). If one extends this notion to quantum mechanics, the associated measure $\mu$ is non-additive. We will explore some properties of this “quantum measure”, its reasonableness in terms of the physics, but its peculiarity on the math side. We will explore the need for a set of properties that can properly characterize the measure and a generalization of the Radon-Nikodym derivative to define a properly extended probability calculus that reduces to the standard additive one on sets of physically distinguishable cases (i.e. orthogonal measurement outcomes).
	
\end{abstract}
\section{Introduction}
It is well-established that quantum probability does not follow classical Kolmogorov probability calculus. Various approaches have been developed to loosen the axioms, of which the use of signed measures is the most successful (e.g. the Wigner quasiprobability distribution).\cite{gleason1957measures,groenewold1946principles,gudder2009quantum,hamhalter2003quantum,monchietti2023measure,moyal1949quantum,sorkin1994quantum,svozil2022extending} On the other hand, the use of non-additive measures (i.e. set functions) has received little attention, which is striking for two reasons. First, we have found that non-additive measures emerge quite naturally in quantum mechanics when looking for a quantum analogue of count of states\cite{aop-phys-QuantumRequiresNonAdditiveMeasures}, which plays a fundamental role in statistical mechanics. Second, there has been an increased use of quantum probability in non-classical decision making (see e.g. Ref. \cite{quantumrev2023}), an area where non-additive measure theory is widely used.

The aim of this article is to introduce the problem for those that work on non-additive set functions but are not familiar with quantum mechanics. The overall goal is to find experts in the field that are interested in collaborating or helping. The general idea is that, in classical mechanics, one is provided with two additive measures, a Liouville measure to quantify states and a measure of probability, and the Radon-Nikodym derivative of the two returns the probability density. The goal is to find a quantum analogue for the two measures and the derivative, thus providing a full ``measure theoretic'' generalization.

\section{Measures in classical mechanics}

In classical mechanics, an ensemble is given by a probability measure $p$ over the Borel algebra of phase space $X$ (i.e. the symplectic manifold charted by the position $q^i$ and momentum $k_i$ of all particles of the system). The space of classical ensembles $\mathcal{E}_C$ (i.e. the space of probability measures over $X$) is a convex space as a convex combination $\lambda p_1 + (1 - \lambda) p_2$ with $\lambda \in [0,1]$ of two probability measures $p_1$ and $p_2$ is also a probability measure.

The space is also equipped with a measure $\mu(U) = \int_U \prod_i dq^i dk_i$ that quantifies states in a volume $U$. Given an ensemble $p$, the probability density $\rho = \frac{dp}{d\mu}$ is the Radon-Nikodym derivative and its entropy is $S(\rho) = - \int_X \rho \log \rho d\mu$.\footnote{The density and entropy must be computed with the Liouville measure $\mu$ or one does not recover the correct physics.} For a uniform distribution over a set $U$, $S(\rho_U) = \log \mu(U)$. Note that we will assume all logarithms to be in base 2, as is customary in information theory.

\section{Finding measures in quantum mechanics}

In quantum mechanics, the space of (pure) states $X$ is the projective space $P(\mathcal{H})$ of a Hilbert space $\mathcal{H}$. An ensemble (or mixed state) is given by a positive semi-definite Hermitian operator $\rho$ of trace one acting on the Hilbert space $\mathcal{H}$ of the system. The details of Hilbert spaces are not essential for our discussion; what is important is that the space of quantum ensembles $\mathcal{E}_Q$ is also a convex space as a convex combination $\lambda \rho_1 + (1 - \lambda) \rho_2$ with $\lambda \in [0,1]$ of two positive semi-definite Hermitian operators $\rho_1$ and $\rho_2$ of trace one is also a positive semi-definite Hermitian operator of trace one. A pure state is an extreme point of the convex set (i.e. an ensemble that cannot be further decomposed into a convex combination of other ensembles) and any ensemble $\rho$ can be expressed as a convex combination of countably many pure states. To each ensemble is associated an entropy $S(\rho) = - tr(\rho \log \rho)$. 

The goal is to characterize a quantum ensemble $\rho$ as a set function $p$ over the Borel algebra of the pure states $X$ in the same way a classical ensemble is a probability measure over phase space. Additionally, we want a set function $\mu$ that quantifies quantum states, similarly to the classical Liouville measure, and to define the equivalent to the probability density $\frac{dp}{d\mu}$. Finally, we want to make sure that classical distributions (i.e. additive measures) are recovered for each quantum observable.\footnote{The idea is that the Borel algebra will contain the lattice of closed subspaces, and therefore we simply need to recover some key results on that sub-lattice.}

\subsection{First attempt for a quantum state-quantifying measure}

As a first attempt\cite{aop-phys-QuantumRequiresNonAdditiveMeasures}, we imposed the same relationship between measure and entropy one has in classical mechanics. That is, given a set $U \subseteq X$ of pure states, we consider the ensemble $\rho_U$ constructed with a uniform convex combination\footnote{The set of pure states $X$ is equipped with a measure invariant under unitary transformation, which can be used to define uniform convex combinations over a Borel set of $X$.}, and we define $\mu(U) = 2^{S(\rho_U)}$.

Since the entropy of a pure state is zero in quantum mechanics, we have $\mu(\{\psi\}) = 1$. For a set of two pure states $U = \{ \psi, \phi \}$, we have 
\begin{equation}
	S(\rho_U) = - \frac{1+\sqrt{p}}{2} \log \frac{1+\sqrt{p}}{2} 
	- \frac{1-\sqrt{p}}{2} \log \frac{1-\sqrt{p}}{2}.
\end{equation}
where $p=|\<\psi | \phi \>|^2$ is given by the square of the inner product, and can be physically understood as the probability of measuring one state given that the other was prepared. Since $0 \leq p \leq 1$, we have $1 \leq S(\rho_U) \leq 2$, which means that $\mu$ is not additive.

Physically, the measure is additive only when the two states are perfectly distinguishable through a measurement or, equivalently, when they can be prepared in the same physical conditions. We can therefore understand $\mu$ as counting states at-all-else-being-equal, and we have arguments as to why this is physically appropriate.

There is, however, an additional property of $\mu$ that make it less appealing: it is non-monotone. That is, the entropy of a uniform distribution over three states can have lower entropy than the uniform distribution of a subset of two. On physical grounds, saying that a bigger set of states has fewer states at-all-else-being-equal does not seem right. On mathematical grounds, as we learned from discussing the topic with experts in non-additive measure theory, generalizations such as the Choquet integral require monotone measures. This led us to revise the approach.

\subsection{Second attempt for quantum measures}

Our second approach, still being finalized\cite{aop-book}, defines $\mu(U) = \sup(2^{S(\hull(U))})$ as the supremum of the entropy reachable with a convex combination of the elements of the set.\footnote{In classical mechanics, this definition recovers $\mu(U) = 2^{S(\rho_U)}$ as the uniform distribution returns the maximum entropy.} If $U$ is a set of one or two pure states, the measure is the same as what we already computed. Therefore the main findings of the previous work carry over. For bigger sets, the measure is monotone because the supremum as well as the hull are monotone functions. Moreover, $\mu$ can be shown to be sub-additive.

Given a mixed state $\rho$, we can define $p_{\rho}(U) = \sup(\{ \lambda \in [0,1] \, | \, \exists \, \rho_1 \in \hull(U), \rho_2 \in \mathcal{E}_Q \; s.t. \;  \rho = \lambda \rho_1 + (1-\lambda) \rho_2 \})$ as the biggest fraction of $\rho$ that can be expressed as a convex combination of $U$. This set function can also be shown to be monotone and sub-additive.

When $\mu$ and $p_\rho$ are restricted to the lattice of closed subspaces, they both become additive on orthogonal subspaces. With this insight we should be able to recover classical probability for quantum measurements.

\section{Open questions}

While we have identified possible candidates for our measures, there are many questions that remain open. First of all, what is the proper notion of derivative, and therefore of integral, that we should be using? Ideally, for every quantum observable $O$ we would like a function $o : X \to \mathbb{R}$ of the pure states so that integrating that function recovers the expectation value for a given mixed state. That is, $tr(\rho O) = \int_X o \frac{dp_\rho}{d\mu} d\mu$. Similarly, we would like to express the entropy as an integral. It is not clear to us whether these features are possible or not.

Additionally, it would be interesting to understand whether these non-additive set functions have any relationship to the sorts of problems for which non-additive set functions are used in decision theory or other areas. We look forward to productive discussions in these areas.

\section{Acknowledgements}

This paper was made possible through the support of grant \#62847 from the John Templeton Foundation. It is part of the ongoing \textit{Assumptions of Physics} project \cite{aop-book}, which aims to identify a handful of physical principles from which the basic laws can be rigorously derived.  We thank Vicen\c{c} Torra, Zuzana Ontkovi\v{c}ov\'{a} and Michel Grabisch for useful pointers and discussion.

%
%
\bibliographystyle{spmpsci}
\bibliography{bibliography}

\begin{thebibliography}{10}
\providecommand{\url}[1]{{#1}}
\providecommand{\urlprefix}{URL }
\expandafter\ifx\csname urlstyle\endcsname\relax
  \providecommand{\doi}[1]{DOI~\discretionary{}{}{}#1}\else
  \providecommand{\doi}{DOI~\discretionary{}{}{}\begingroup
  \urlstyle{rm}\Url}\fi

\bibitem{aop-book}
Carcassi, G., Aidala, C.A.: Assumptions of Physics.
\newblock Michigan Publishing (2021).
\newblock \doi{10.3998/mpub.12204707}.
\newblock \urlprefix\url{https://assumptionsofphysics.org/book/}

\bibitem{aop-phys-QuantumRequiresNonAdditiveMeasures}
Carcassi, G., Aidala, C.A.: How quantum mechanics requires non-additive
  measures.
\newblock Entropy \textbf{25}(12) (2023).
\newblock \doi{10.3390/e25121670}

\bibitem{gleason1957measures}
Gleason, A.M.: Measures on the closed subspaces of a {H}ilbert space.
\newblock Journal of Mathematics and Mechanics \textbf{6}(6), 885 (1957)

\bibitem{groenewold1946principles}
Groenewold, H.J.: On the principles of elementary quantum mechanics.
\newblock Springer (1946)

\bibitem{gudder2009quantum}
Gudder, S.: Quantum measure and integration theory.
\newblock Journal of Mathematical Physics \textbf{50}(12) (2009)

\bibitem{hamhalter2003quantum}
Hamhalter, J.: Quantum measure theory, vol. 134.
\newblock Springer Dordrecht (2003).
\newblock \doi{https://doi.org/10.1007/978-94-017-0119-8}

\bibitem{monchietti2023measure}
Monchietti, E., Massri, C., de~Barros, J.A., Holik, F.: Measure-theoretic
  approach to negative probabilities.
\newblock arXiv preprint arXiv:2302.00118  (2023)

\bibitem{moyal1949quantum}
Moyal, J.E.: Quantum mechanics as a statistical theory.
\newblock In: Mathematical Proceedings of the Cambridge Philosophical Society,
  vol.~45, pp. 99--124. Cambridge University Press (1949)

\bibitem{quantumrev2023}
Plotnitsky, A., Haven, E.: The Quantum-Like Revolution, 1st edn.
\newblock Springer Cham (2023)

\bibitem{sorkin1994quantum}
Sorkin, R.D.: Quantum mechanics as quantum measure theory.
\newblock Modern Physics Letters A \textbf{9}(33), 3119--3127 (1994)

\bibitem{svozil2022extending}
Svozil, K.: Extending {K}olmogorov’s axioms for a generalized probability
  theory on collections of contexts.
\newblock Entropy \textbf{24}(9), 1285 (2022)

\end{thebibliography}
\end{document}